\documentclass[twocolumn]{aastex631}

\shorttitle{Toward a live homogeneous database of solar active regions}
\shortauthors{Ruihui Wang et al.}

\graphicspath{{./}{figures/}}
\usepackage{multirow}
\usepackage{booktabs}
\usepackage{hyperref}
\usepackage{amsmath}
\usepackage{threeparttable}
\usepackage{verbatim}

\begin{document}

\title{Toward a live homogeneous database of solar active regions based on SOHO/MDI and SDO/HMI synoptic magnetograms. \uppercase\expandafter{\romannumeral2}. parameters for solar cycle variability}

\correspondingauthor{Jie Jiang}
\email{jiejiang@buaa.edu.cn}

\author{Ruihui Wang}
\affiliation{ School of Space and Environment, Beihang University, Beijing, People’s Republic of China; jiejiang@buaa.edu.cn}
\affiliation{Key Laboratory of Space Environment Monitoring and Information Processing of MIIT, Beijing, People’s Republic of China}

\author{Jie Jiang}
\affiliation{ School of Space and Environment, Beihang University, Beijing, People’s Republic of China; jiejiang@buaa.edu.cn}
\affiliation{Key Laboratory of Space Environment Monitoring and Information Processing of MIIT, Beijing, People’s Republic of China}

\author{Yukun Luo}
\affiliation{ School of Space and Environment, Beihang University, Beijing, People’s Republic of China; jiejiang@buaa.edu.cn}
\affiliation{Key Laboratory of Space Environment Monitoring and Information Processing of MIIT, Beijing, People’s Republic of China}

\begin{abstract}
Solar active regions (ARs) determine solar polar fields and cause solar cycle variability within the framework of the Babcock-Leighton (BL) dynamo. The contribution of an AR to the polar field is measured by its dipole field, which results from flux emergence and subsequent flux transport over the solar surface. The dipole fields contributed by an AR before and after the flux transport are referred to as the initial and final dipole fields, respectively. For a better understanding and prediction of solar cycles, in this paper, we provide a database including AR’s initial and final dipole fields and the corresponding results of their bipolar magnetic region (BMR) approximation from 1996 onwards. We also identify the repeated ARs and provide the optimized transport parameters. Based on our database, we find that although the commonly used BMR approximation performs well for the initial dipole field, it exhibits a significant deviation for the final dipole field. To accurately assess an AR's contribution to the polar field, the final dipole field with its real configuration should be applied. Despite the notable contributions of a few rogue ARs, approximately the top 500 ARs ordered by their final dipole fields are necessary to derive the polar field at the cycle minimum. While flux transport may increase or decrease the dipole field for an individual AR, its collective impact over all ARs in a cycle is a reduction in their total dipole field.

\end{abstract}


\section{Introduction} \label{sec:intro}
Solar active regions (ARs) are common features on the solar surface \citep{Van2015}, originating from the emergence of flux tubes \citep{Fan2009, Cheung2010}. They determine solar polar fields and cause the solar cycle variability. In the Babcock-Leighton (BL) dynamo, the emergence of ARs and the subsequent transport of erupted AR flux on the surface reverse the existing poloidal field, generating a new poloidal field with the opposite polarity \citep{Babcock1961, Leighton1964}. This poloidal field is then stretched by differential rotation, giving rise to the toroidal field, which influences the strength of the subsequent solar cycle. Therefore, a correlation between the poloidal field, or polar field, during solar minimum and the strength of the subsequent cycle is expected, and observations confirm the linear correlation between them \citep{Schatten1978, Jiang2007, Munoz-Jaramillo2013}. If the contribution of all ARs in a given cycle is known, we can predict the polar field in solar minimum, and the strength of the following cycle. Such predictions are crucial for forecasting the frequency and intensity of future solar activities, as well as for planning future space missions \citep{Jiang2018, Petrovay2020SCpd}.

The contribution of an AR to the polar field can be quantified by its axial dipole field, also known as the axial dipole moment. The axial dipole field of an AR originates from flux emergence and changes due to flux transport across the solar surface \citep{Wang1991}. The dipole field of an AR immediately after its emergence is termed the initial dipole field ($D_i$). The initial dipole field is primarily influenced by the latitude difference between the two polarities of ARs, described by the polarity separation and tilt angle when optimized to a bipolar magnetic region (BMR). 

However, $D_i$ only describes the contribution of flux emergence. Subsequent to emergence, AR dipole fields increase or decrease during flux transport \citep{Wang1991, Jiang2014apj}. The dipole field after flux transport is referred to as the final dipole field ($D_f$). During flux transport, a portion of the AR flux diffuses across the equator due to supergranular diffusion and is then transported to the pole via meridional flow and supergranular diffusion \citep{Jiang2014ssr, WangYM2017, Yeates2023}. Simultaneously, an equivalent amount of net flux with the opposite polarity to the cross-equatorial flux is transported to the other pole. The cross-equatorial flux dictates the final flux distribution of ARs and, consequently, determines $D_f$ \citep{Cameron2013, Petrovay2020AM}. An AR's latitude and transport parameters, which include supergranular diffusivity and meridional flow velocity, influence the cross-equatorial flux, leading to the variation in $D_f$.

For ARs emerging near the solar equator, the absolute values of their $D_i$ are dramatically increased by flux transport. Given that these ARs may not strictly obey Hale's law and Joy's law, they can significantly increase or decrease the polar field and the strength of the following cycle. Those ARs are regarded as rogue ARs. \cite{Jiang2015} demonstrated that the weak cycle 24 is related to several rogue ARs in cycle 23. Moreover, \cite{Nagy2017} found that some rogue ARs have the potential to halt the dynamo, leading to a grand minimum. The contribution of rogue ARs can only be accurately assessed through $D_f$.

ARs are usually approximated as symmetric bipolar magnetic regions (BMRs) in studies concerning their emergence and evolution. However, the actual configuration of an AR is complex and often deviates significantly from a BMR. Firstly, the two polarities of an AR are commonly asymmetric in terms of both flux and area \citep{Driel-Gesztelyi1990, Tlatov2014}. This asymmetry can substantially reduce or even reverse the dipole field of an AR compared to a symmetric one \citep{Iijima2019, Wang2021}. Secondly, there are many ARs with complex configurations, such as $\delta$-type ARs. The dipole fields of these ARs sometimes change their sign during flux transport \citep{Jiang2019}, which can not be described by symmetric BMRs. \cite{2020Yeates} demonstrated that the BMR approximation tends to overestimate the solar axial dipole field during the minimum compared to real ARs. Including the real configuration of ARs, \cite{Wang2021} proposed a general method that calculates $D_f$ of ARs more accurately than the method based on BMR approximation. They both emphasize the significance of considering the real configuration of ARs in obtaining their $D_f$.

The initial dipole field, $D_i$, specifically characterizes the contribution of flux emergence to the polar field, while $D_f$ encompasses both the contributions of flux emergence and flux transport. When considering the limitations of BMR approximation, the real contribution of an AR to the polar field can only be accurately measured by its $D_f$ while taking into account its real configuration. As flux transport can either increase or decrease the dipole fields of individual ARs, a key question arises: what is the overall impact of flux transport on all ARs in a solar cycle? \cite{Wang1991} shows that the overall effect of flux transport over a solar cycle is a decrease in the dipole field originating from flux emergence. Their study only includes ARs in cycle 21, necessitating further research.

A database providing both $D_i$ and $D_f$ contributes to investigating the roles of flux emergence and flux transport in the BL mechanism. Moreover, such a database can serve as a valuable resource for studying the influence of ARs, particularly rogue ones, on the polar field and solar cycle variability, known as dynamo effectivity \citep{Petrovay2020AM}, and subsequently aid in predicting solar cycles.

However, most existing AR databases only provide basic parameters and parameters related to space weather \citep{WangYM1989, Bobra2014, SMARPsASHARPs, MunozJaramillo2021, Sreedevi2023}. An exception is the work by \cite{2020Yeates}, which provides a database containing $D_i$ and $D_f$ for cycle 24 based on SHARPs data \citep{Bobra2014}. Here, we aim to construct a live, comprehensive, and homogeneous AR database covering cycles 23, 24, and 25 in a series of papers. This database will encompass two sets of parameters: the first set includes fundamental AR parameters like number, area, and flux, while the second set includes parameters directly for solar cycle variability, such as $D_i$, $D_f$, and their BMR approximations, $D_i^B$ and $D_f^B$. In the first paper of the series \citep{Wang2023}, we develop an automated AR detection method based on SOHO/MDI and SDO/HMI synoptic magnetograms \citep{MDI, HMI}. We then provide the first set of parameters based on these detections. In this paper, we first address the issue of repeated ARs in the detections and then present the second set of parameters. Based on a substantial number of AR observations in our database, we systematically explore the disparities between $D_i$ and $D_f$ and show the impact of flux transport and flux emergence on the dipole field. Furthermore, we analyze the effects of BMR approximation, as well as the influence of a few rogue ARs and numerous ARs with small $D_f$.
 
This paper is organized as follows. In Section \ref{sec:repeat ARs}, we identify and remove repeat ARs. In Section \ref{sec:Calculation DF}, we optimize the transport parameters and calculate the $D_i$, $D_f$, and their BMR approximations. In Section \ref{sec:stastics}, we compare $D_i$ and $D_f$ of real ARs, show the impact of BMR approximation on $D_i$ and $D_f$, and analyze the impact of many ARs with small $D_f$ on the cumulative final dipole field. In Section \ref{sec:conclusion}, we summarize and discuss the above results.

\section{Identification and removal of repeat active regions} \label{sec:repeat ARs}

Some ARs with strong fluxes can live for many Carrington rotations (CRs) \citep{Harvey1993, Li2008}. Their impacts on the end-of-cycle polar field are usually significant. The removal of these repeated observations is crucial for an accurate reconstruction of the polar field. In the first paper of the series, \cite{Wang2023}, we have developed an AR automatic detection method, which consists of five modules. The first module employs adaptive intensity threshold segmentation, while the second module incorporates morphological closing and opening operations. These modules jointly eliminate background magnetic fields and decayed ARs and extract kernel pixels of ARs. The third module, region growing, utilizes these kernel pixels as seeds to recover all pixels associated with each AR. Modules 4 and 5 work together to eliminate decayed ARs that persist even after the region-growing process. The fourth module involves a morphological closing operation and a small-region removal operation. The fifth module consists of a morphological dilation operation for merging neighboring regions and a unipolar region removal operation based on their flux imbalance. 

\begin{figure}[htbp!]
\centering
\includegraphics[scale=0.36]{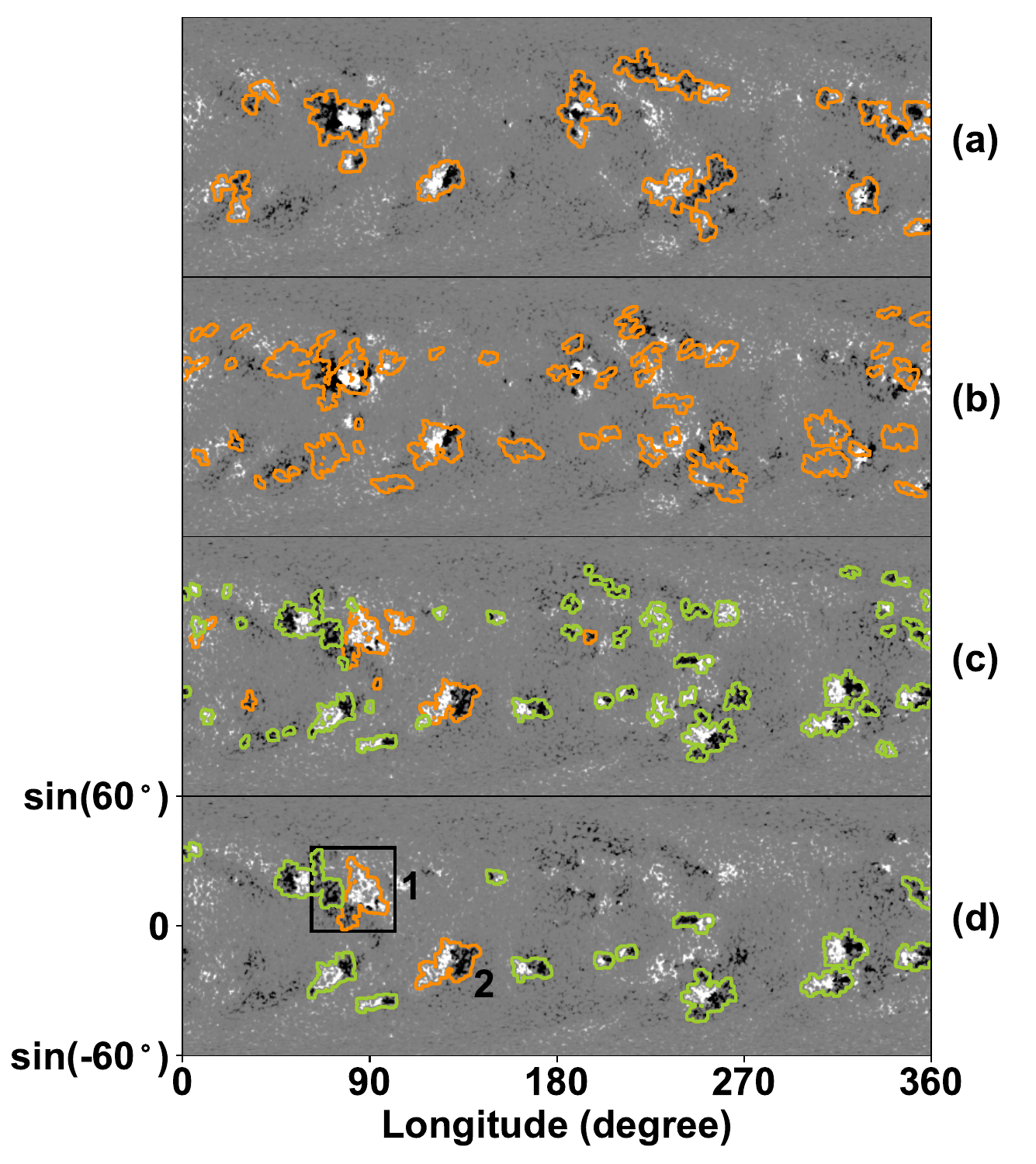}
\caption{Illustration of the repeat-AR-removal module, using the synoptic magnetogram of CR 1968 as an example. Panel (a) is the map of CR 1967, with the AR detection result outlined in orange lines. Panel (b) is the map of CR 1967, overplotted with the masks of ARs detected in CR 1968. Panel (c) is the AR detection result of CR 1968 after the repeat-AR-removal module, and panel (d) is the final detection result after module 5. In panels (c) and (d), the repeat ARs are labeled with orange lines, and new ARs are labeled with green lines. In panel (d), two detected repeat ARs are labeled with No. 1 and 2. The black square labels the corresponding result, estimated by the naked eye, of the repeat AR No.1.
\label{fig1}}
\end{figure}

The aforementioned detection method processes each synoptic magnetogram independently and thus can not identify repeated ARs. To address this limitation, we introduce a repeat-AR-removal module positioned between Module 4 and Module 5. Adding this module before Module 5 aims to avoid the potential impact of merging neighbor regions in Module 5, enhancing the method's ability to identify and remove repeated ARs. The step-by-step procedures of the repeat-AR-removal module are visually presented in Figure \ref{fig1}, with the CR 1968 map serving as an example.

In the process of removing repeat ARs in the CR 1968 map, we also utilize the synoptic magnetogram from the last CR, 1967. After Module 4, the AR detection algorithm has identified all potential ARs and eliminated small, decayed AR segments. Subsequently, we derotate the detected ARs to the last CR to compensate for the effects of differential rotation. Then we obtain masks. Each mask of an AR, depicted by orange lines in panel (b), labels the region where the flux of the AR in CR 1968 comes. We apply the differential rotation profile  
\begin{equation}
  \omega(\lambda) = \omega_a+\omega_bsin^2(\lambda)+\omega_csin^4(\lambda),  
\end{equation}
where $\lambda$ is the latitude. $\omega_a=0.362^{\circ}day^{-1}$, $\omega_b=-2.040^{\circ}day^{-1}$, and $\omega_c=-1.487^{\circ}day^{-1}$.  These values are provided by the header file of the MDI and HMI synoptic magnetograms.

Then we conduct a comparison between the unsigned flux in CR 1968 ($F_{n}^{us}$, where n denotes the n-th map, i.e., CR 1968 map) and the unsigned flux of the CR 1967 map within the mask ($F_{n-1}^{us}$, where n-1 denotes the n-1-th map, i.e., CR 1967 map) for each detected AR in CR 1968. If the ratio $F_{n-1}^{us}/F_{n}^{us}$ exceeds a defined threshold ($T_{fr}$) and the AR satisfies the polarity condition, we regard the AR as a repeat observation of an AR in CR 1967. Considering the impact of other flows on the solar surface, such as the meridional flow and supergranulation, the mask obtained by derotating an AR to the last CR may not precisely represent the true area where the flux of the AR originates. Therefore, the threshold doesn't necessarily have to be set to 1, and after experiments, we set $T_{fr}=0.7$ for optimal results. The polarity condition is 
\begin{equation}
    \left | \frac{F_n}{F_{n}^{us}} \frac{F_{n-1}}{F_{n-1}^{us}} \right | sign(\frac{F_n}{F_{n-1}}) > -0.5,
\end{equation}
where $F_n, F_{n-1}$ are the net flux of AR in CR 1968 and the net flux of CR 1967 map within the mask, respectively. The term $sign(\frac{F_n}{F_{n-1}})$ is utilized to determine whether the sign of the AR net flux in two maps is the same. The ratios $\frac{F_n}{F_{n}^{us}}$ and $\frac{F_{n-1}}{F_{n-1}^{us}}$ are employed to assess the flux imbalance of AR in CR 1968 map and AR in CR 1967 map, respectively. This condition prevents the unipolar regions, which differ in polarity from the unipolar regions in the last map, from being wrongly identified as repeat regions. The result is shown in panel (c), where the region labeled in the orange line represents repeat ARs that have been successfully detected and removed. Finally, we apply Module 5 to merge neighbor regions and remove unipolar regions. The final detection result of CR 1968 is shown in panel (d). 

The repeat-AR-removal module successfully removes two repeat ARs in the MDI map of CR 1968, labeled with No. 1 and 2. A visual comparison between panels (d) and (a) reveals that AR No. 2 is a clear repeat observation of a former AR. However, the situation for AR No. 1 is more complex. It constitutes only a portion of the entire AR, where a new AR emerges near the former one. Such ARs are commonly referred to as activity complexes (ACs) \citep{Harvey1993, Gaizauskas1983, Wang2020}. Comparing the two maps in CRs 1967 and 1968 with the naked eye, we can find that the repeat regions in the AC should be regions within the black square. While the actual repeat region might be slightly larger than the region we detected, the module still identifies the repeat AR within the AC. The results of the CR 1968 map demonstrate that our repeat-AR-removal module is not only capable of removing obvious repeat ARs but can also address cases of repeat decayed ARs within ACs. It's worth noting that, at times, the module may face challenges in accurately detecting repeat regions due to the segmentation of single ARs in modules 1-4 of \cite{Wang2023}. 

After removing repeat ARs, the total number of detected ARs decreases from 2579 to 2495 over two solar cycles. Specifically, the number of ARs in cycles 23 and 24 reduces from 1481 to 1436 and from 1098 to 1059, respectively. \cite{2020Yeates} detects 143 repeat ARs during cycle 24, a higher count compared to our detection result of 39 repeat ARs. Discrepancies in the number of repeat ARs could be attributed to the use of a smaller threshold, $T_{fr}=0.7$, in our detection in contrast to the threshold $T_{fr}=1$ employed by \cite{2020Yeates}. Additionally, discrepancies may also arise from the fact that some repeat ARs in our results are just part of the whole AC and do not affect the AR number. 

\section{Calculation of dipole fields} \label{sec:Calculation DF}

\subsection{Calculation methods of dipole fields} \label{sus:Cal methods}

\begin{figure}[htbp!]
\centering
\includegraphics[scale=0.33]{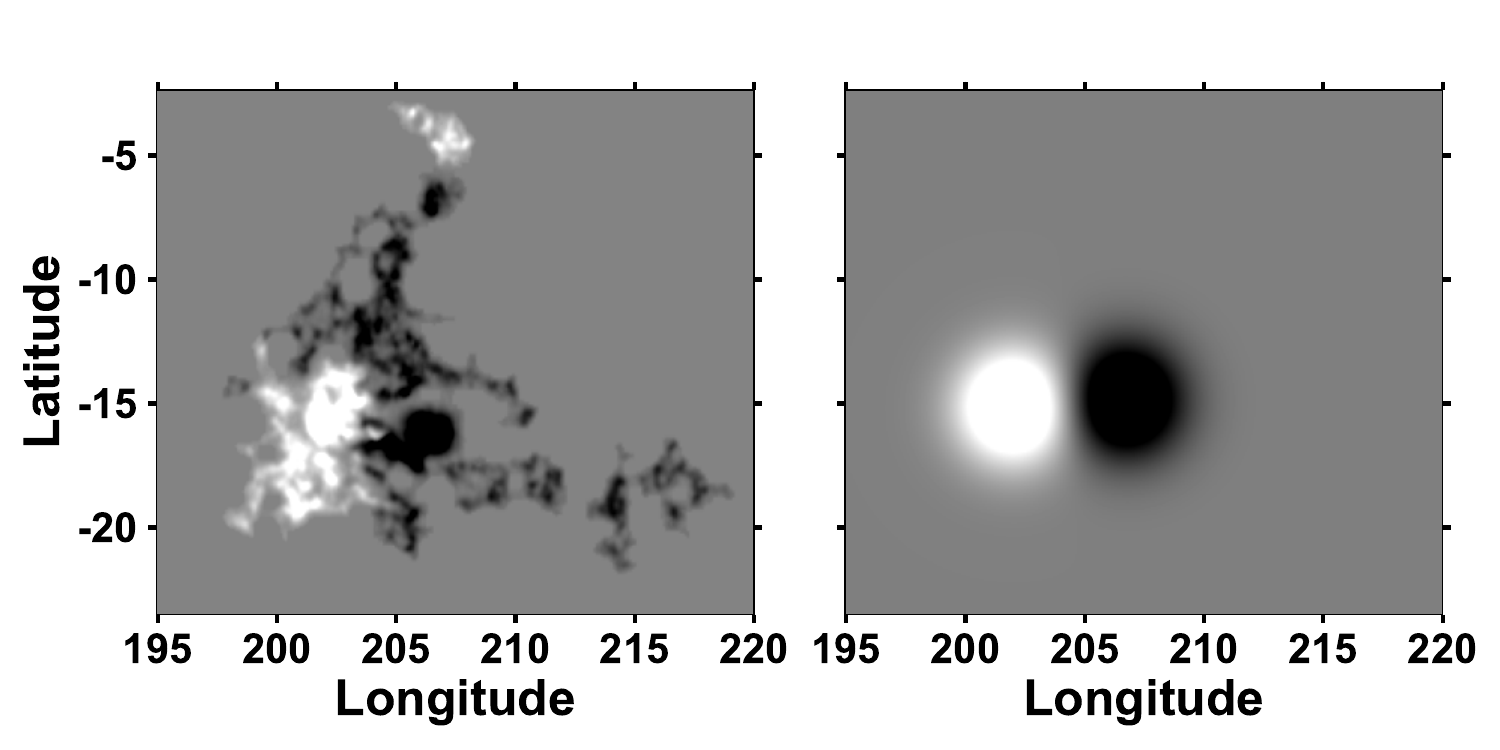}
\caption{Illustration of an AR (left panel) and its corresponding BMR approximation (right panel). The AR is detected in the MDI synoptic magnetogram of CR 1961, with label No. 6 in our database. Its NOAA number is 08936.
\label{fig:ARBMRcase}}
\end{figure}

After removing repeat ARs in our database, we calculate the axial dipole fields of ARs, including the initial dipole field ($D_i$) and final dipole field ($D_f$). Each parameter is determined using two distinct methods: one based on the actual configuration of ARs, termed the AR method, and the other based on the BMR approximation, referred to as the BMR method. For the BMR approximation, we approximate each polarity of AR as a finite-sized Gaussian-like polarity patch \citep{van_Ballegooijen1998, Jiang2014ssr}. Figure \ref{fig:ARBMRcase} shows an example of an AR and its BMR approximation.

The axial dipole field of an AR is the coefficient of the $Y^0_1$ term in the spherical harmonic expansion of its radial magnetic fields. Mathematically, this is expressed as:
\begin{equation}
D_i=\frac{3}{4\pi} \iint B(\lambda ,\varphi )sin\lambda cos\lambda d\lambda d\varphi,
\label{eqDi}
\end{equation}
where $\lambda$ and $\varphi$ denote latitude and longitude, respectively, and $B(\lambda ,\varphi )$ is the radial magnetic field distribution of the AR. When applying the BMR approximation to an AR, Equation (\ref{eqDi}) can be simplified as:
\begin{equation}
    D_i^{B}=\frac{3}{4\pi R_{\odot}^2}\Phi d_{\lambda} cos\lambda ,
\label{eqDiB}    
\end{equation}
where $\Phi$ represents the unsigned flux of one polarity of the BMR. $d_{\lambda}$ is the latitude difference of the BMR. It can be calculated using the latitude of the positive polarity minus the latitude of the negative polarity, or as $d_{\lambda} = d \sin \alpha$, where $d$ is the polarity separation, and $\alpha$ is the tilt angle \citep{Wang1991, Yeates2023}.

Equations (\ref{eqDi}) and (\ref{eqDiB}) are used to quantify the axial dipole field of an AR when it emerges on the solar surface, referred to as the initial dipole field ($D_i$). After a BMR emergence, its axial dipole field changes due to meridional flow and supergranular diffusion. The ratio ($f_\infty$) between the dipole field at the end of a BMR's evolution (final dipole field, $D_f$) and $D_i$ follows a Gaussian relation about the BMR emergence latitude $\lambda_0$. The relation is expressed as $f_\infty \propto \exp({-a \lambda_ 0^2})$, where $a$ is a constant \citep{Jiang2014apj}. \cite{Petrovay2020AM} provide an algebraic expression for $f_\infty$, and thus provide an algebraic method to calculate the final dipole field of a BMR ($D_f^{B}$). The method is 
\begin{equation}
    D_f^{B}=f_\infty D_i^B,
    \label{eqDfB}
\end{equation}
where $f_\infty$ is given by
\begin{equation}
    f_\infty=\frac{a}{\lambda_{R}^B} exp{\left (-\frac{\lambda _0^2}{2(\lambda _{R}^B)^2} \right )}, a=\frac{n+1}{n+2} \sqrt{\frac{2}{\pi}}.
    \label{eqfinf}
\end{equation}
In Equation (\ref{eqfinf}), $\lambda _0$ represents the emergence latitude of the BMR, and $n$ describes the approximated polar field profile in solar minimum, i.e. $B(\lambda) \propto \sin^n(\lambda)$. The value of $n$ is set to 7 in our calculations, following the value of \cite{WangYM2009}. $\lambda _{R}^B$ is the dynamo effectivity range, calculated by:
\begin{equation}
    \lambda _{R}^B =\sqrt{\frac{\eta }{R_{\odot}^2\Delta u} + \sigma_0^2}.
    \label{eqLRB}
\end{equation}
Here, $\eta $ is the supergranular diffusivity, $\Delta u$ is the divergence of the meridional flow at the equator, i.e., $\frac{1}{R_{\odot}} \left.\frac{\mathrm{d} u}{\mathrm{d} \lambda } \right|_{\lambda=0} $, and $\sigma_0$ is the initial Gaussian width of the BMR.

However, Equation (\ref{eqDfB}) just applies to BMRs. For ARs with arbitrary configurations, \cite{Wang2021} provides a more general method expressed as
\begin{equation}
D_f=A_0\iint B(\lambda,\varphi ) erf(\frac{\lambda }{\sqrt{2}\lambda _R })cos\lambda d\lambda d\varphi,    
\label{eqDf}
\end{equation}
where $A_0$ is a constant, set to 0.21 according to \cite{Wang2021}. In this equation, the dynamo effectivity range $\lambda _R$ is defined as
\begin{equation}
    \lambda _R =\sqrt{\frac{\eta }{R_{\odot}^2\Delta u}}.
\end{equation}
This is different from $\lambda _{R}^B$ in Equation (\ref{eqfinf}), for the BMR method. The form of Equation (\ref{eqDf}) has been slightly adjusted from that presented in \cite{Wang2021} to improve its conciseness and clarity.

\subsection{Optimization of the transport parameter $\lambda_R$} \label{sus:optimzt TP}
\begin{figure}[htbp!]
\centering
\includegraphics[scale=0.35]{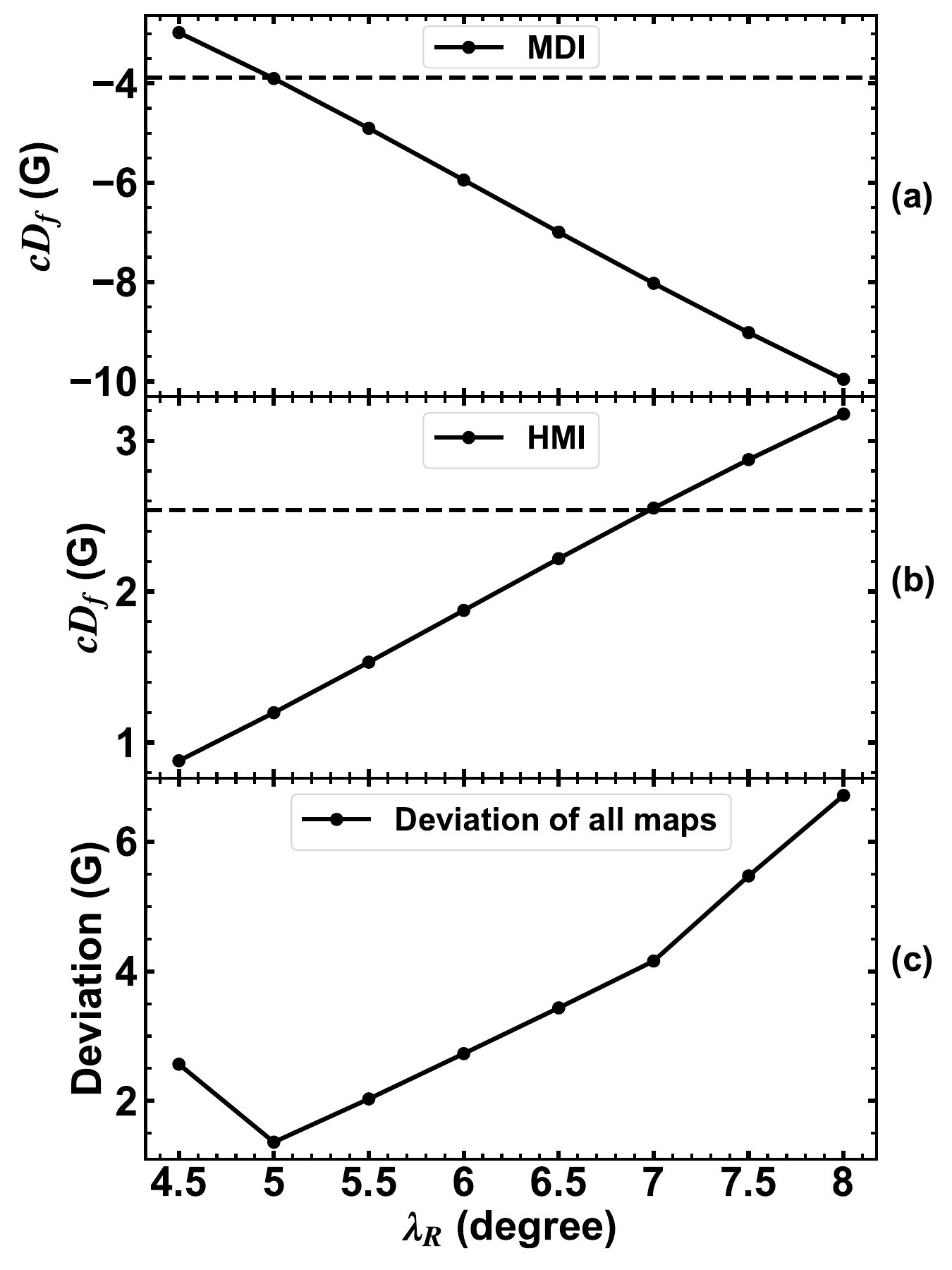}
\caption{Comparison between cumulative final dipole fields ($cD_f$) for different transport parameters ($\lambda_R$) and dipole field observations of MDI maps (CRs 1909-2096) and HMI maps (CRs 2097-2225). Panels (a) and (b) depict the results of MDI and HMI maps, respectively. Black dots represent the calculated $cD_f$, and dashed lines show dipole field observations. Panel (c) illustrates the total deviation between $cD_f$ and dipole field observations across all synoptic magnetograms.
\label{fig:optim_lamdr}}
\end{figure}

In the computation of final dipole fields ($D_f$), the term $\frac{\eta }{R_{\odot}^2\Delta u}$ is present in both the BMR method and the AR method. The meridional flow (\textit{u}) and supergranular diffusivity ($\eta$) are less constrained by observations \citep{Petrovay2020AM, Jiang2023}. A comprehensive discussion on the observations of $\Delta u$ and $\eta$ is provided in \cite{Jiang2023}. $\Delta u$ and $\eta$ are sometimes optimized with magnetogram observations when employed in Surface Flux Transport (SFT) models \citep{Lemerle2015, Whitbread2017}. According to the BL mechanism, the solar axial dipole field is built up through the emergence and evolution of ARs \citep{Babcock1961, Leighton1964}. Consequently, the cumulative final dipole fields ($cD_f$) in one solar cycle, defined as the sum of all ARs' $D_f$, should be equal to the difference between the axial dipole fields at the end-of-cycle solar minimum and that at the start-of-cycle minimum. Since the AR method is considered more accurate than the BMR method for ARs with real configurations \citep{Wang2021}, we utilize the $D_f$ calculated by the AR method to optimize $\lambda_R$, which is equivalent to optimizing $\frac{\eta }{R_{\odot}^2\Delta u}$. We also refer to $\lambda_R$ as the transport parameter.

We conduct a comparison between the $cD_f$ and the solar dipole fields obtained from the MDI and HMI synoptic magnetograms. The MDI maps cover the time range from CR 1909 to CR 2096, roughly corresponding to cycle 23, while the HMI maps cover the time range from CR 2097 to CR 2225, roughly corresponding to cycle 24. To calculate the solar dipole field, we use the synoptic magnetograms with the polar field corrected. The correction is performed by interpolation using relatively well-observed data during September/March for many years \citep{Sun2011, sun2018,Luo2023}. The dipole field for the beginning or the end of a cycle is determined by averaging the dipole fields calculated in six CRs near this time.

The comparison is depicted in Figure \ref{fig:optim_lamdr}. As $\lambda _R$ increases, the absolute values of $cD_f$ for both the MDI and HMI maps also increase. The optimal value of $\lambda _R$ for the MDI maps is $5^\circ$, while for the HMI maps, it is $7^\circ$. The transport parameter $\lambda _R$ used in the SFT simulation based on HMI observations by \cite{2020Yeates} is $6.35^\circ$, which is only slightly smaller than the parameter we use. The minor discrepancy may be attributed to the differences in the magnetic field distributions of ARs in our database and that of \cite{2020Yeates}.

To minimize the overall deviation between the $cD_f$ and the solar dipole fields obtained from both the MDI and HMI maps, the optimal $\lambda _R$ is also $5^\circ$. This value is not the smallest among the transport parameters used in past research, such as $4^\circ$ in \cite{Lemerle2015} and $4.36^\circ$ in \cite{WangYM2017}. The total deviation for the optimal $\lambda _R$ is still significant. Factors contributing to this deviation may include the detection of new ARs in the ACs, limitations in synoptic magnetogram observations, and limitations in polar field observations. Although our detection method, incorporating the repeat-AR-removal module, can eliminate some repeated regions and detect new ARs in the ACs, it may still miss some repeat regions or erroneously remove newly emerged flux. Synoptic magnetograms lack observations of the far-side solar disk and have relatively low time resolution. The limitations in synoptic magnetogram observations may result in not all detected ARs being in their fully emerged phase, leading to the potential loss of some emerged AR flux. Additionally, synoptic maps fail to capture some short-lived ARs, particularly those emerging on the far-side of the sun. Considering the small tilt angle of the Sun’s rotation axis relative to the ecliptic and the projection effect of magnetic field observations, measurements of the polar magnetic field are poor. Magnetic field observations near polar regions often exhibit a high noise level and contain missing values. Interpolation is applied to obtain missing data \citep{Sun2011, Luo2023}, thereby affecting the polar field. As a result, the solar dipole field from the synoptic maps may also deviate from the actual value.

We use $\lambda _R = 5^\circ$ to calculate $D_f$, and meanwhile, we compare the results of $D_f$ calculated with $\lambda _R = 7^\circ$ to show the impact of $\lambda _R$. In addition, we also provide the processing programs along with the database and anyone can calculate the parameter $D_f$ with another $\lambda _R$ if necessary.

\subsection{Calculation of dipole fields based on the detected ARs}
With the optimal transport parameter, $\lambda_R = 5^\circ$, we apply the calculation methods described in Section \ref{sus:Cal methods} to AR detection results to calculate dipole fields for each AR. First, we balance the magnetic flux of each AR by increasing the flux of the weak polarity because a part of the flux of the more dispersed polarity, usually the following polarity, is missed by the detection. The method is the same as \cite{Wang2020} and \cite{Jiang2019}. Then, we get the magnetic field distribution, $B(sin\lambda, \varphi)$, used for the AR method, directly from AR detection of the maps. Flux $\Phi$, latitude difference $d_\lambda$, and initial Gaussian width $\sigma_0$ used in the BMR method are also calculated from the AR detection results.

\begin{figure}[htbp!]
\centering
\includegraphics[scale=0.37]{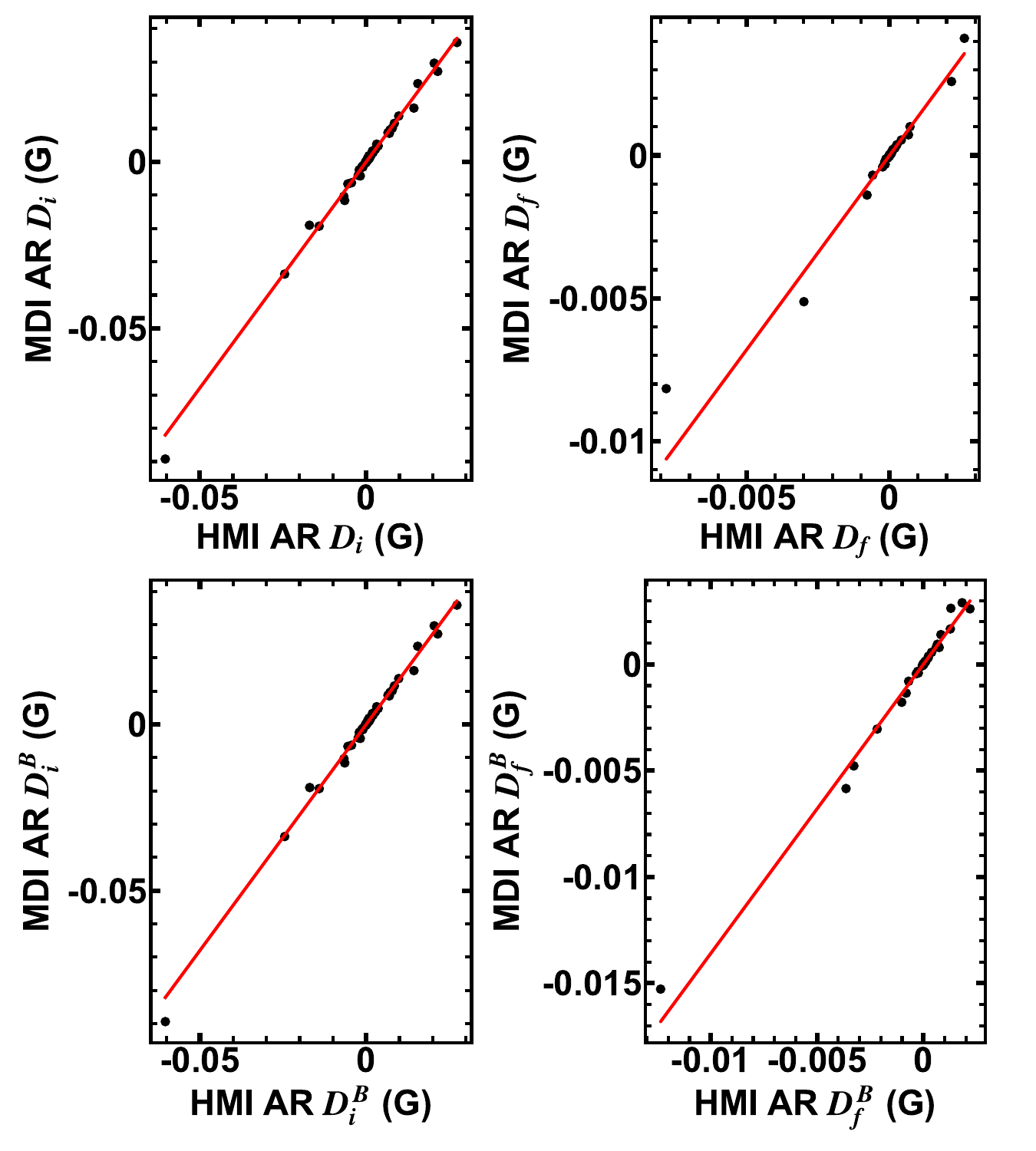}
\caption{Calibration of AR initial dipole fields ($D_i$, $D_i^B$) and final dipole fields ($D_f$, $D_f^B$) from MDI and HMI synoptic magnetograms. The red line in each panel indicates the calibration based on AR flux, with a slope of 1.36.
\label{fig4}}
\end{figure}

To align with the results from MDI maps, the AR flux detected in HMI synoptic magnetograms needs to be multiplied by a factor of 1.36 \citep{Wang2023}. Since equations of both AR and BMR methods are linear in terms of magnetic fields, we also calibrate the AR dipole fields in HMI maps by applying a factor of 1.36. Comparing the parameters of the same AR detected in MDI and HMI maps during the overlap period, as shown in Figure \ref{fig4}, the factor works well for all parameters.

\section{statistical properties of dipole fields}\label{sec:stastics}
After removing repeat ARs in our detection and calculating dipole fields, we update the detected ARs in our database and provide the second set of parameters. There are 2892 ARs in our database now, from CR 1909 to CR 2278 (1996.5 - 2023.11). The second set of parameters includes the initial dipole field ($D_i$), the final dipole field ($D_f$), and the BMR approximation of the two parameters, $D_i^B$ and $D_f^B$. Together with the first set of parameters, i.e. AR basic parameters, such as location, area, and flux, our database can now provide 17 parameters. The second set of parameters can describe the contribution of ARs to the solar polar field and contribute to the understanding and prediction of solar cycles. To show the properties of the second set of parameters, we compare the $D_i$ and $D_f$, analyze the impact of BMR approximation on the dipole fields, and investigate the impact of ARs with small $D_f$ on the cumulative final dipole fields.

\subsection{Difference between $D_i$ and $D_f$} \label{sus:DiDf}
\begin{figure}[htbp!]
\centering
\includegraphics[scale=0.33]{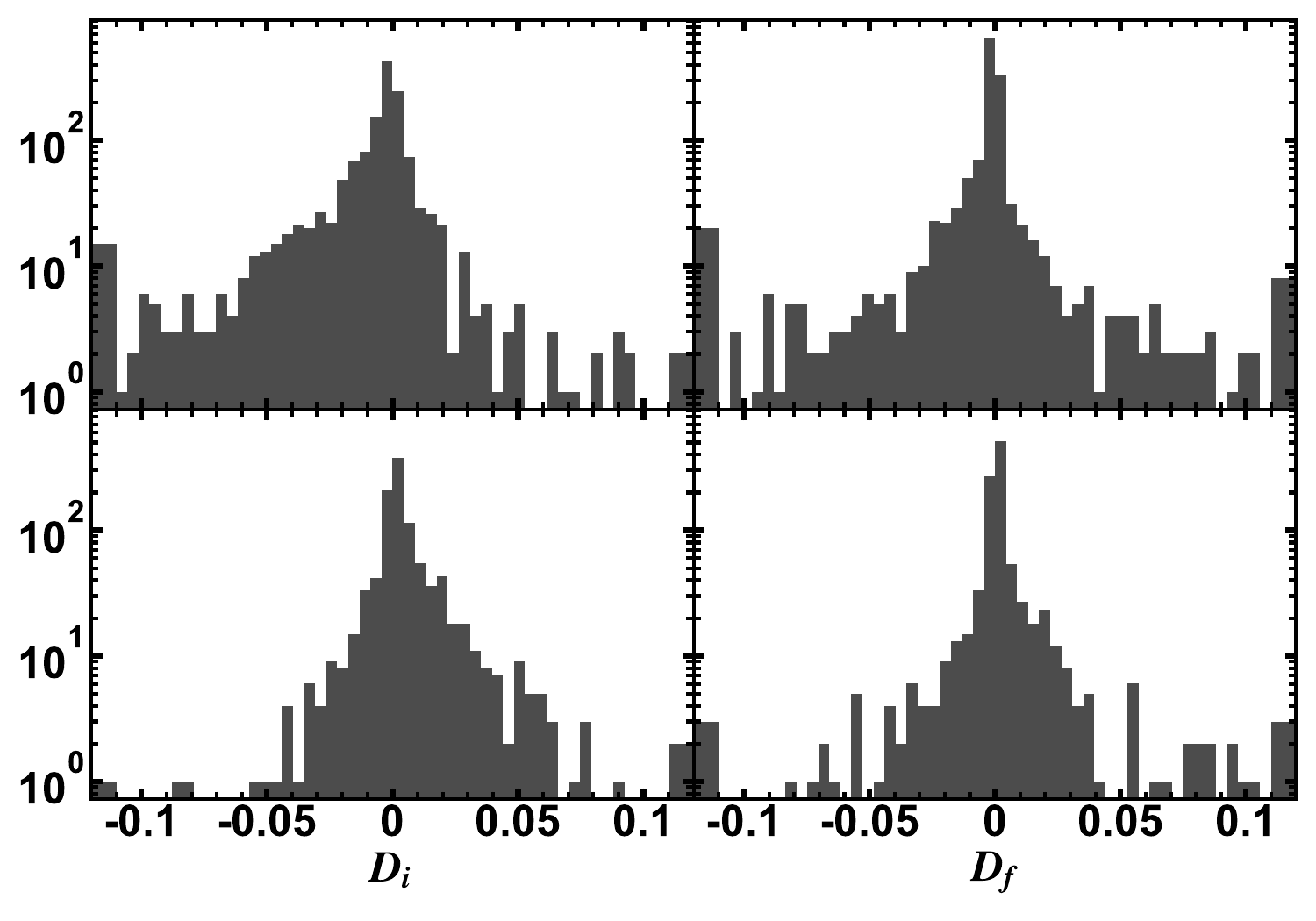}
\caption{Histograms of initial dipole field ($D_i$) and final dipole field ($D_f$) for cycles 23 (top panels) and 24 (bottom panels). Lower (-0.11 G) and upper (0.11 G) thresholds are applied in each panel. Data outside these thresholds are combined and shown in the first and last bins.
\label{fig:distrb}}
\end{figure}

First, we systematically compare the initial dipole field ($D_i$) and final dipole field ($D_f$) of ARs with real configurations to illustrate the impact of flux transport on each AR and all ARs in a solar cycle. Figure \ref{fig:distrb} shows the distribution of $D_i$ and $D_f$. Both $D_i$ and $D_f$ exhibit concentration around zero, with the distribution of $D_f$ being more centered than $D_i$. $D_i$ tends to be negative during cycle 23 and positive in cycle 24, while $D_f$ shows a more symmetric distribution around zero for both cycles. Although some ARs exhibit significant $D_i$ and $D_f$, the majority have relatively small $D_i$ or $D_f$. In two cycles, 1709 ARs, approximately 68\% of all ARs, have an absolute $D_i$ smaller than 0.01 G, which is less than 1\% of the total dipole field. The corresponding number for $D_f$ is 2002, accounting for about 80\%. While certain ARs can notably influence the solar dipole field, the impact of most ARs is limited.

\begin{figure}[htbp!]
\centering
\includegraphics[scale=0.43]{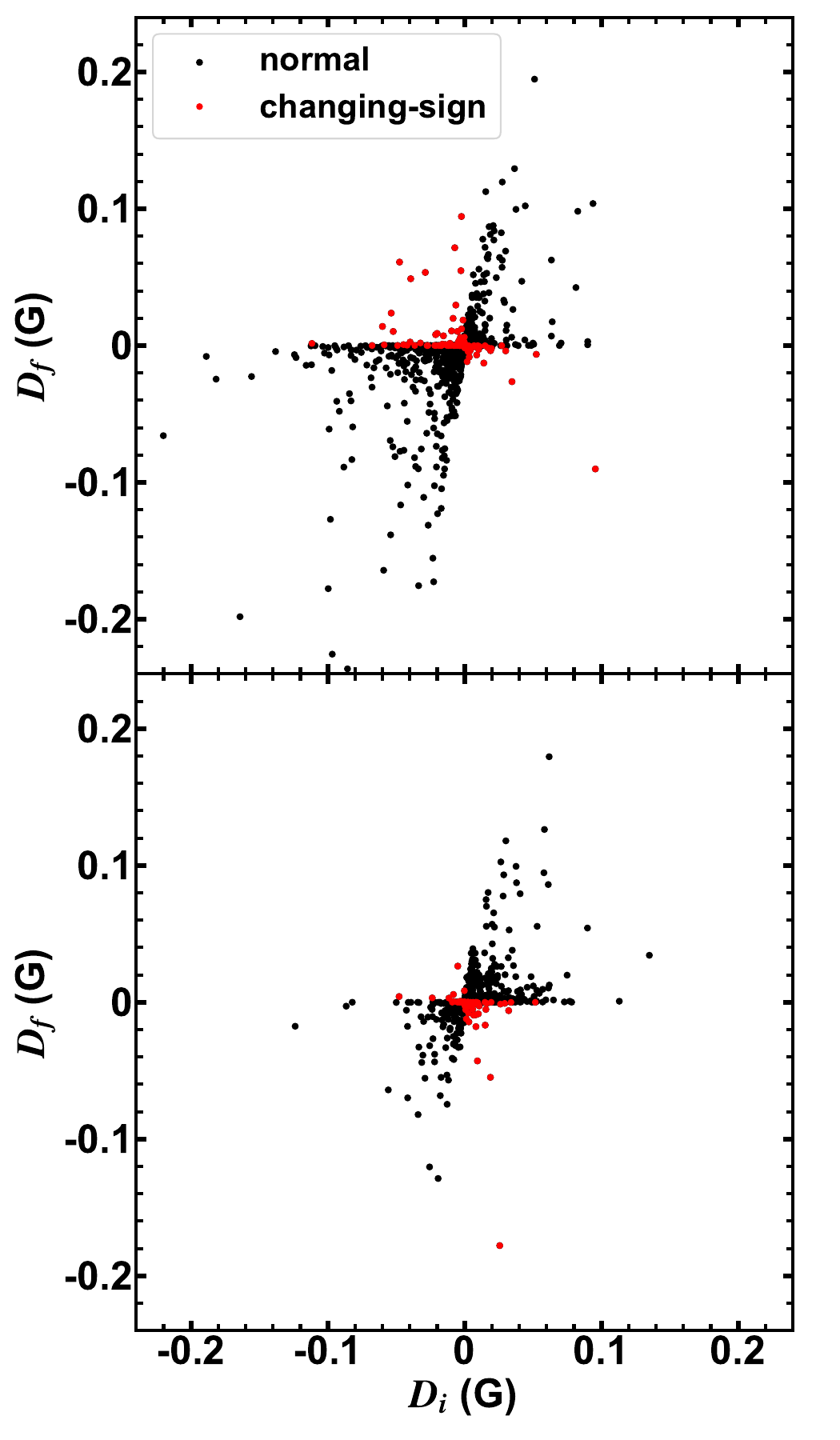}
\caption{Comparison between the initial dipole field ($D_i$) and the final dipole field ($D_f$) of cycles 23 (top panel) and 24 (bottom panel). Data of nine ARs in cycle 23 with absolute $D_f$ larger than 0.24 G are not shown. The black dots label the ARs whose $D_i$ and $D_f$ are of the same sign (normal ARs), and the red dots label the ARs whose $D_i$ and $D_f$ are of different sign (changing-sign ARs). One AR among the nine ARs not shown here changes the sign of the dipole field. 
\label{fig:DiDf}}
\end{figure}

Figure \ref{fig:DiDf} shows the comparison between $D_i$ and $D_f$. For some ARs, $D_f$ is larger than $D_i$, smaller than $D_i$, or even exhibits a different sign (referred to as changing-sign ARs), as denoted by the red dots. Figure \ref{fig:ARBMRcase} is an example of changing-sign ARs, with $D_i$ of -0.0064 G and $D_f$ of 0.0296 G. The changing-sign ARs are first reported by \cite{Jiang2019}. They will be further discussed in Section \ref{sus:BMR approximation}. Among ARs with the same sign for $D_f$ and $D_i$, there are 898 ARs for cycle 23 and 687 ARs for cycle 24 whose axial dipole field decreases (absolute value of $D_f$ smaller than that of $D_i$). Conversely, there are 403 and 281 ARs for cycles 23 and 24, respectively, whose axial dipole field increases. These changes in dipole fields contribute to the difference between the distributions of $D_f$ and $D_i$, as illustrated in Figure \ref{fig:distrb}. As the axial dipole fields of most ARs decrease, the distribution of $D_f$ is more concentrated around zero compared to $D_i$. In both cycles, the number of $D_f$ in bins close to zero surpasses that of $D_i$. Additionally, as flux transport increases the dipole field of certain ARs, the instances with absolute value greater than 0.1 G for $D_f$ are also more than those for $D_i$, as shown in Figure \ref{fig:distrb}.

When $\lambda_R$ is increased from $5^\circ$ to $7^\circ$, the number of ARs with $D_f$ smaller than 0.01G decreases to 1809, accounting for approximately 72.5\% of all ARs. In cycles 23 and 24, the number of ARs with dipole field increasing rises from 684 to 991, and for ARs with dipole field decreasing reduces from 1585 to 1374. The contribution of most ARs to the solar dipole field is still small and the number of ARs with dipole fields decreasing still exceeds those with dipole fields increasing. The impact of $\lambda_R$ is limited.

As for the cumulative dipole fields, defined as the sum of all dipole fields, the absolute value of the cumulative final dipole field ($cD_f$) in each cycle is smaller than that of the cumulative initial dipole field ($cD_i$). The values of $cD_f$ are -3.91 G for cycle 23 and 1.63 G for cycle 24, while the values of $cD_i$ are -11.85 G for cycle 23 and 4.70 G for cycle 24. However, the $cD_f$ is influenced by the transport parameter $\lambda_R$, increasing with $\lambda_R$. Consequently, $cD_f$ might surpass $cD_i$ for a larger $\lambda_R$. Nevertheless, $\lambda_R$ is optimized to minimize the deviation between $cD_f$ and the solar axial dipole field of each cycle. The dipole field observations of MDI and HMI are -3.88 G and 2.54 G, respectively and the corresponding $cD_i$ values for the MDI and HMI time range are -11.69 G and 3.34 G. These indicate that the absolute values of $cD_i$ are also greater than those of the axial dipole fields of cycles 23 and 24. Therefore, $cD_f$ will consistently be smaller than $cD_i$ as long as $\lambda_R$ is appropriately chosen. The result is consistent with the finding of cycle 21 by \cite{Wang1991}. The fact that the absolute values of $cD_f$ are smaller than that of $cD_i$ indicates that the solar axial dipole field primarily originates from flux emergence. The net effect of flux transport is to reduce the solar dipole field that is obtained during flux emergence. In the jargon of the BL-type mechanism, the generation of the poloidal flux ($\alpha$-effect) mainly happens in the flux emergence, while the surface flux transport mainly decreases rather than generates the poloidal flux.

\subsection{Impact of the BMR approximation on dipole fields}\label{sus:BMR approximation}

\begin{figure}[htbp!]
\centering
\includegraphics[scale=0.4]{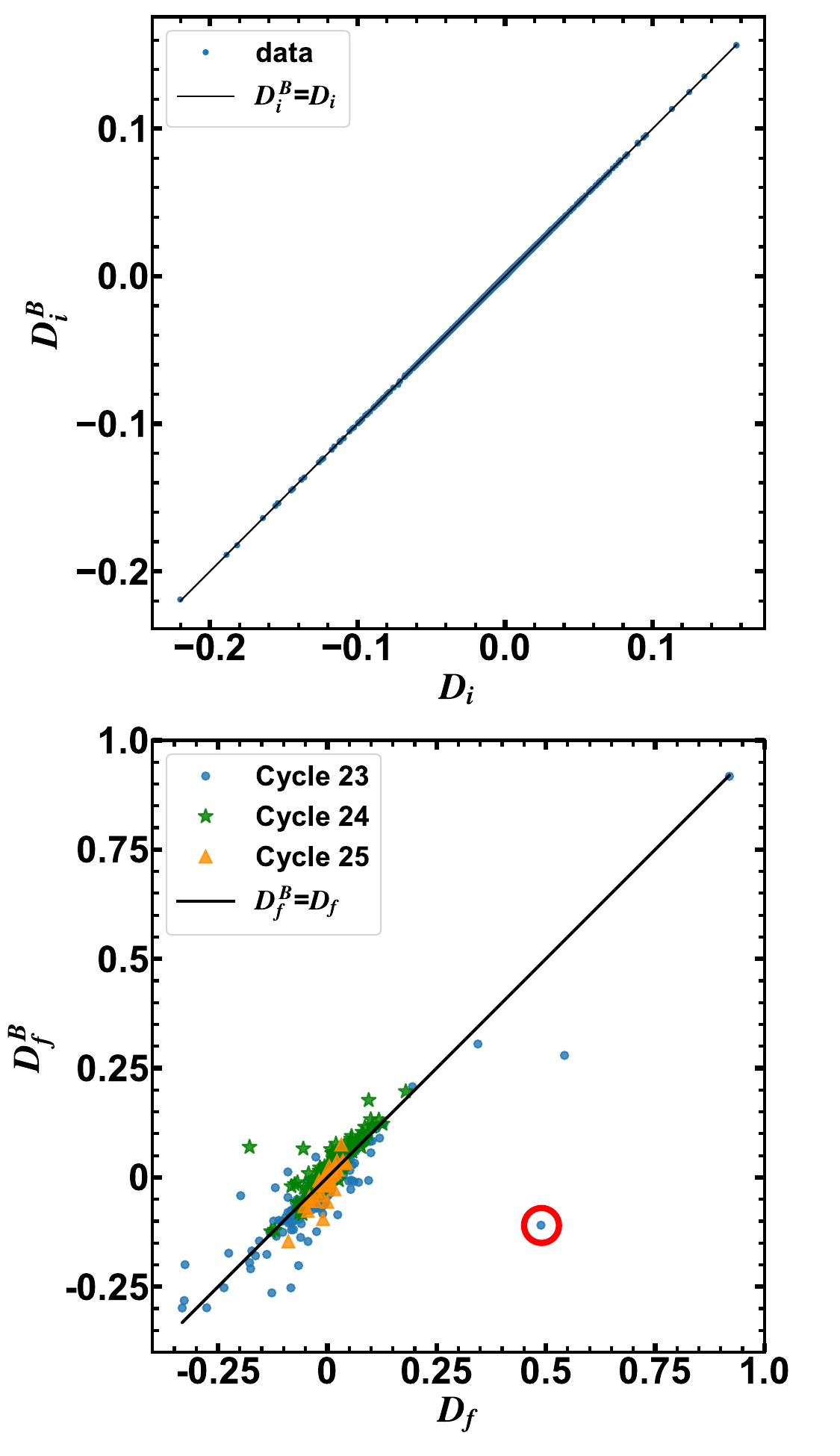}
\caption{Comparisons between dipole fields. Top panel: the initial dipole field calculated using the AR method ($D_i$) and the BMR method ($D_i^{B}$). Bottom panel: the final dipole field calculated using the AR method ($D_f$) and the BMR method ($D_f^B$). The lines in both panels represent the values calculated by the two methods are the same. The red circle in the bottom panel labels an AR with a significant difference between $D_f^B$ and $D_f$.
\label{fig:ARBMR}}
\end{figure}

The BMR approximation of AR is widely used in AR emergence and evolution research. Here we compare the initial dipole field ($D_i$) and final dipole field ($D_f$) calculated using the AR method with those calculated using the BMR method ($D_i^{B}$ and $D_f^{B}$) to test the performance of the BMR approximation. The results are shown in Figure \ref{fig:ARBMR}. For the initial dipole field, the values of the BMR approximation ($D_i^B$) are highly consistent with that of the AR method ($D_i$). This finding is consistent with the research of \cite{2020Yeates}. Furthermore, the cumulative initial dipole field of BMR approximation ($cD_i^B$) and that of AR ($cD_i$) are also nearly the same. For example, $cD_i^B$ is -11.854 G in cycle 23, while $cD_i$ is -11.850 G. It shows that the BMR approximation of the initial dipole field proposed by \cite{Wang1991} (Equation (\ref{eqDiB})), is in line with the AR method (Equation (\ref{eqDi})). The BMR approximation is more efficient than the AR method and is therefore recommended for use.

However, the real contribution of ARs to the polar field at the solar minimum is $D_f$, not $D_i$. For $D_f$, the BMR approximation yields less satisfactory results. The result aligns with the findings of \cite{Wang2020} but with more data than it. Although the deviations between $D_f^B$ and $D_f$ for many ARs are small, there are instances where the deviations are significant, exceeding 0.1 G, a magnitude comparable to the $D_f$ of these ARs. 

Additionally, the BMR approximation sometimes incorrectly predicts the sign of the final dipole field for certain ARs. Given that $D_i$ is nearly equal to $D_i^B$, and the signs of $D_f^B$ and $D_i^B$ are the same (as seen from Equations (\ref{eqDfB} - \ref{eqfinf})), the sign of $D_f^B$ aligns with $D_i$. Consequently, ARs with different signs of $D_f$ and $D_f^B$ also exhibit different signs of $D_f$ and $D_i$. These ARs are referred to as changing-sign ARs in Section \ref{sus:DiDf}. Among 2952 ARs in cycles 23, 24, and part of 25, there are 266 changing-sign ARs in total. Similarly, \cite{2020Yeates} identifies 53 ARs out of 1090 ARs where $D_f$ obtained with the BMR method and the AR method have different signs. The change in sign is typically attributed to the complex configurations of these ARs, which significantly differ from the BMR model \citep{Jiang2019}. For instance, the AR shown in Figure \ref{fig:ARBMRcase} changes the sign of the dipole field due to the small section of positive polarity located at about $-5^\circ$ latitude, which is ignored by BMR approximation. Just like the small positive polarity, these regions emerging near the equator can induce significant variations in both the AR's dipole field and that of the Sun. They should be precisely considered in research, not be approximated.

Some changing-sign ARs have a notable impact on the solar polar field; for example, $D_f$ for the AR labeled with a red circle in Figure \ref{fig:ARBMR} is 0.49 G, whereas the BMR approximation $D_f^B$ of it is -0.11 G. The deviation between the BMR approximation and the AR method of the AR is about 0.6 G, about 15\% of the total dipole field of cycle 23. To accurately predict the contribution of these changing-sign ARs to the polar field, it is recommended to employ their final dipole field ($D_f$) while considering their real configurations. The BMR approximation is prone to incorrectly predicting the dipole field of these ARs.

Moreover, the cumulative values of $D_f^B$ and $D_f$ exhibit more significant differences. The $cD_f^B$ and $cD_f$ values are -7.95 G and -3.91 G for cycle 23, and 3.85 G and 1.63 G for cycle 24. The $cD_f^B$ value for each cycle is approximately twice the $cD_f$ value. As the BMR approximation displays a substantial deviation from the AR method, predictions relying on the BMR approximation, as exemplified by \cite{Pal2023}, may be affected by this deviation, raising concerns about their reliability. For the BMR approximation, $D_f^B/D_i^B$ is proportional to $exp{\left (-\frac{\lambda _0^2}{2(\lambda _{R}^B)^2} \right )}$, as shown in Equation (\ref{eqfinf}), where $\lambda_{R}^B =\sqrt{\frac{\eta }{R_{\odot}^2\Delta u} + \sigma_0^2}$. Here, $\sigma_0$ represents the initial Gaussian width of the BMR. In our analysis, we take half the polarity separation as $\sigma_0$ for the BMR. However, if we set $\sigma_0$ as $0^\circ$, the cumulative values of $D_f^B$ become more similar to those of $D_f$, resulting in -4.39 G for cycle 23 and 2.56 G for cycle 24. The impact of $\sigma_0$ on the final dipole field is also discussed by \cite{Iijima2019}. The $\sigma_0$ of BMRs influences their $D_f^B$ but is challenging to determine accurately. The AR method proposed by \cite{Wang2021} (Equation (\ref{eqDf})), which does not require an initial Gaussian width $\sigma_0$, can calculate the AR final dipole field more accurately. This shows the advantage of the AR method.

When $\lambda_R = 7^\circ$, the number of changing-sign ARs decreases to 150. The corresponding $D_f$ and $D_f^B$ for the AR labeled in Figure \ref{fig:ARBMR} become 0.33 G and -0.08 G, respectively. There is still a large deviation of 0.41 G, approximately 8\% of the total dipole field. Significant deviation persists between $cD_f^B$ and $cD_f$. For instance, $cD_f$ for cycle 23 is -8.05 G, while $cD_f^B$ is -11.12 G. Adjusting the initial Gaussian width $\sigma_0$ to 0 improves the BMR approximation, yielding $cD_f^B$ of -8.55 G, closer to $cD_f$. When $\lambda_R = 7^\circ$, although the changing-sign ARs are relatively strongly influenced, other results only weakly change. The BMR approximation remains poor for the final dipole field, and setting $\sigma_0$ to 0 works better than employing half the polarity separation.

In summary, although the BMR approximation performs well for the initial dipole field, it yields inferior results for the final dipole field, which describes the true contribution of an AR to the polar field in the solar minimum. When measuring an AR's contribution to the polar field, its final dipole field should be utilized while taking into account its real configuration, i.e., the parameter $D_f$ in our database.

\subsection{Impact of ARs with small $D_f$ on the cumulative final dipole field} \label{sus:cDf}

\begin{figure}[htbp!]
\centering
\includegraphics[scale=0.36]{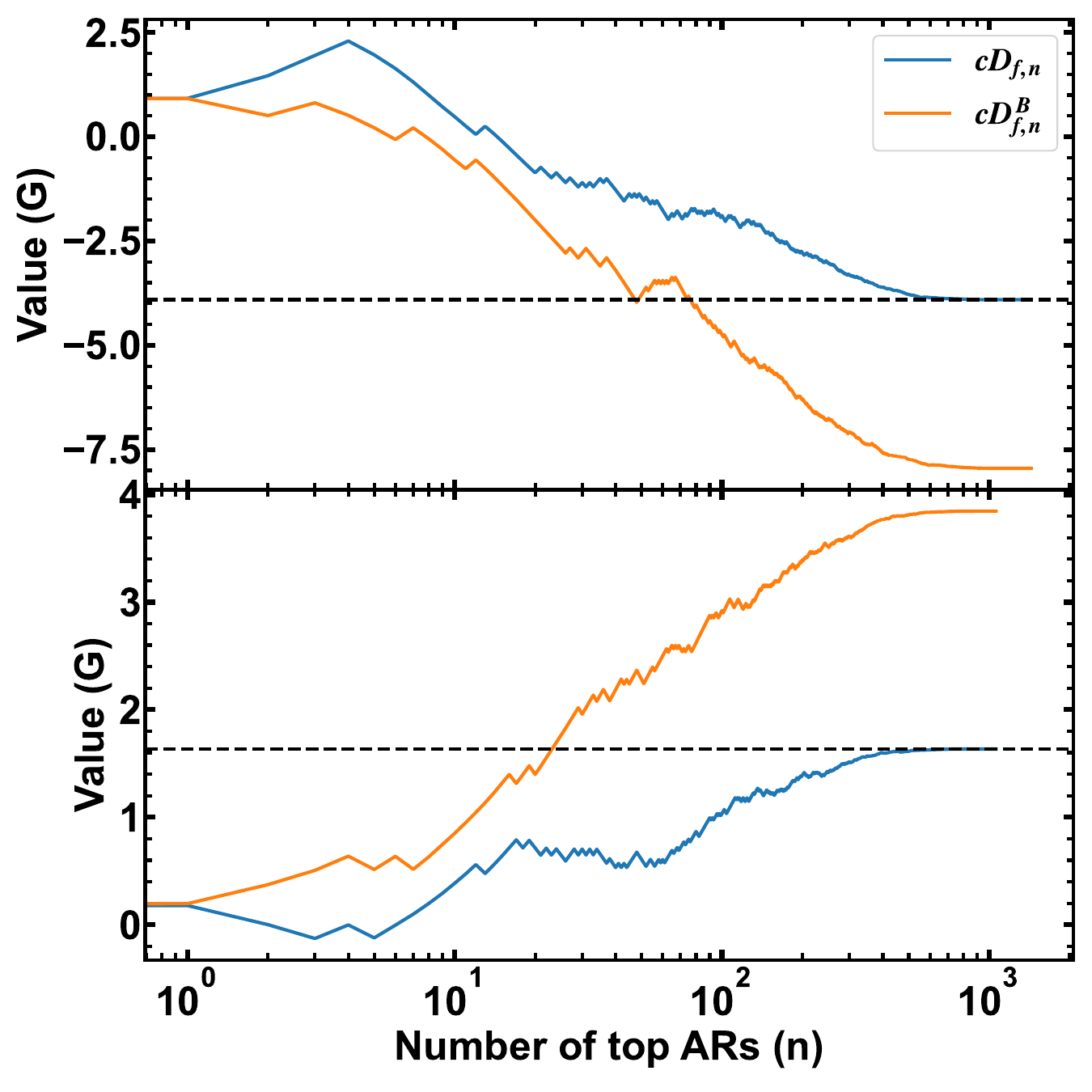}
\caption{Variation of the cumulative final dipole field for different numbers of top ARs ($cD_{f,n}$, $cD_{f,n}^B$). ARs are ranked by the absolute value of their final dipole field. Both $cD_{f,n}$ for the AR method (blue line) and $cD_{f,n}^B$ for the BMR method (orange line) are shown. The top panel is the result of cycle 23 and the bottom panel is for cycle 24. The black dashed lines in two panels represent the cumulative final dipole fields of all ARs ($cD_f$) in each cycle.
\label{fig7}}
\end{figure}

As shown in Section \ref{sus:DiDf}, the $D_f$ of most ARs are small. To research the impact of numerous ARs with small $D_f$, we rank ARs according to the absolute values of their $D_f$ and calculate the cumulative final dipole field for different numbers of top ARs ($cD_{f,n}$). The analysis is conducted with the AR method ($cD_{f,n}$) and the BMR method ($cD_{f,n}^B$). The variations of $cD_{f,n}$ and $cD_{f,n}^B$ with the number of ARs (n) are presented in Figure \ref{fig7}. For $cD_{f,n}$ calculated with the AR method, its deviation with the cumulative final dipole field of all ARs ($cD_f$) in each cycle firstly increases due to certain rogue ARs with contributions opposite to those of other ARs. Then, the deviation gradually decreases under the effect of plenty of ARs with small $D_f$. When the AR number increases to about 500, the $cD_{f,n}$ of two cycles both becomes almost equal to the $cD_f$. The deviations are 3.3\% for cycle 23 and 1.2\% for cycle 24. If a similar deviation threshold is set for cycle 24, the AR number decreases to about 350, roughly one-third of all ARs (1059). The 500 ARs also constitute about one-third of all ARs in cycle 23, during which the total number of ARs is 1436. The smallest absolute values of $D_f$ in the top one-third ARs are 0.0028 G and 0.0023 G for cycles 23 and 24, respectively. 

For $cD_{f,n}^B$ calculated with the BMR method, the orange line in Figure \ref{fig7}, its deviations with the cumulative final dipole field for all BMRs, $cD_f^B$, gradually decreases for each cycle with the increase of BMR number without the initial increase for the deviation between $cD_{f,n}$ and $cD_f$. It shows that these ARs with opposite contributions, which cause the initial increase for the deviation between $cD_{f,n}$ and $cD_f$, are largely a result of their complex configurations that significantly differ from the BMR. As the number of BMRs increases to around 500, the $cD_{f,n}^B$ values for cycles 23 and 24 also become nearly equivalent to the $cD_f^B$. 

When $\lambda_R = 7^\circ$, the initial increase in the variation of deviation between $cD_{f,n}$ and $cD_f$ with the number of ARs (n) disappears in cycle 24 but persists in cycle 23. This further supports the presence of rogue ARs in cycle 23. The deviations between $cD_{f,n}$ of the top 500 ARs and $cD_{f}$ are 5.1\% and 3\% for cycles 23 and 24, respectively. When applying the top 350 ARs, the top one-third ARs of cycle 24, the deviation for cycle 24 is 8\%. Although these values are larger than those for $\lambda_R = 5^\circ$, they are still deemed acceptable. The top 500 ARs, or the top one-third ARs, remain sufficient for reconstructing the axial dipole fields during the solar minimum. As for $cD_{f,n}^B$, there is no significant change.

Our results align with the findings of \cite{Nagy2020} and \cite{Whitbread2018}, but our study requires a smaller number of ARs, approximately 500 ARs compared to their 750 ARs. These studies reveal that while several rogue ARs can induce noticeable variations in the solar polar field, the collective impact of numerous ARs with smaller contributions to the polar field cannot be disregarded. Notably, not all ARs are essential, and approximately the top 500 ARs, or the top one-third of all ARs, are sufficient for building up the polar field during the solar minimum. \cite{Hofer2024} also emphasizes the importance of small ARs on the polar field variation.

\section{Conclusion and Discussion} \label{sec:conclusion}
In this paper, we have provided AR parameters including the initial dipole field ($D_i$), the final dipole field ($D_f$), and the Bipolar Magnetic Region (BMR) approximations for both parameters, denoted as $D_i^B$ and $D_f^B$. These AR parameters are part of our live homogeneous database of solar ARs since 1996 and are relevant to solar cycle variability, aiming to improve our understanding and prediction of solar cycles. The properties of these parameters have also been investigated.

Since some ARs can live for many CRs and significantly impact the end-of-cycle polar field, we first introduce a repeat-AR-removal module to identify and remove the repeat detections in our database. This module is not only capable of removing obvious repeat ARs but can also address cases of repeat decayed ARs within activity complexes (ACs). Then we optimize the transport parameter $\lambda_R$ because it significantly affects the final dipole field, but is poorly constrained by observations. We compare the cumulative final dipole field ($cD_f$) obtained with different $\lambda_R$ values and the solar dipole fields derived from the MDI and HMI synoptic magnetograms. It reveals that the optimal $\lambda_R$ to minimize the deviation between $cD_f$ and solar dipole field is $5^\circ$. 

Following the removal of repeat ARs and the optimization of $\lambda_R$, we calculate the dipole fields and update the time range of our AR database. Our database now comprises 2892 ARs spanning from CR 1909 to CR 2278 (1996.5 - 2023.11). The database and associated codes are freely available to everyone on GitHub\footnote{\texttt{AR database:} \url{https://github.com/Wang-Ruihui/A-live-homogeneous-database-of-solar-active-regions/tree/with-repeat-AR-Removal-and-params-dipole-fields}.} and version 2.0 is archived in Zenodo \citep{wang_2024_11115058}. In exploring the properties of the dipole fields, we analyze the distinctions between $D_i$ and $D_f$, assess the impact of BMR approximation on the dipole field, and evaluate the influence of ARs with small $D_f$.

The analysis reveals that $D_f$ of an individual AR may be smaller, larger, or even with a different sign compared to its corresponding $D_i$. However, the cumulative final dipole field ($cD_f$) of all ARs during a cycle is notably smaller than $cD_i$ for cycles 23 and 24. It aligns with the findings for cycle 21 as reported by \cite{Wang1991}. The cumulative final dipole field $cD_f$ smaller than $cD_i$ suggests that flux emergence is the primary source of the solar axial dipole field in the BL mechanism, and the net effect of flux transport is to decrease the solar dipole field initially obtained during flux emergence. Hence the surface flux transport must be also considered properly to obtain a realistic BL dynamo \citep{Cameron2012, Jiang2013, Zhang2022}. It is noticeable that the conclusion is based on data from three cycles, and further validation with data from more cycles would enhance its robustness.

While the BMR approximation works well for $D_i$, it exhibits poor performance for $D_f$. The deviation between the BMR approximation and real ARs is even more pronounced for $cD_f$. Specifically, for ARs that change the sign of their dipole fields during flux transport, the BMR approximation fails to predict the sign of their $D_f$. It introduces a considerable deviation compared to $D_f$ without approximation. Therefore, it is advisable to utilize the final dipole field while taking into account the real configuration of ARs, i.e. $D_f$ in our database, to measure an AR's contribution to the polar field accurately.

While the impact of most ARs is limited, their effects can not be ignored. About the top 500 ARs, ordered by their absolute $D_f$, are necessary to build up the polar field in the solar minimum. This result is consistent with the results of \cite{Whitbread2018} and \cite{Nagy2020}. 

In optimizing transport parameter $\lambda_R$, the deviation between $cD_f$ and solar dipole field for the optimal $\lambda_R$ still can not be ignored, suggesting that $5^\circ$ may not be so perfect. The deviation may be caused by the detection of new ARs in the ACs, limitations in synoptic magnetogram observations, and limitations in polar field observations. The deviation may also be caused by the possible variation of transport parameter $\lambda_R$ with solar cycles \citep{WangYM2009, Whitbread2017}. To assess the impact of $\lambda_R$ on the presented results, we also provide the results for $\lambda_R = 7^\circ$. These results indicate that while $\lambda_R$ significantly influences $D_f$, the distinctions between $D_i$ and $D_f$, the impact of BMR approximation on $D_f$, and the influence of ARs with small $D_f$ are only weakly affected by $\lambda_R$.

This is the second paper of a series that aims to construct a live, comprehensive, and homogeneous AR database. In the first paper of the series \citep{Wang2023}, we identified all ARs based on the MDI and HMI synoptic magnetograms and provided some fundamental parameters as the first set of parameters. In this paper, we provide the dipole fields as the second set of parameters. In the future, we will continue to provide more parameters and deeply analyze AR properties, particularly those of rogue ARs. An important parameter to be explored is the Degree of Rogueness (DoR). \cite{Nagy2020} suggests that only about the top 10-50 DoRs are required to build up the polar field in solar minimum, while the other ARs could be simplified. Their results significantly reduce the number of ARs necessary and show the importance of the DoR. However, based on our database, we find DoR is affected by the statistical result applied in its calculation, and the result of the top 50 DoRs while simplifying the other ARs may be not so ideal as shown by \cite{Nagy2020}. Further exploration and discussion on the DoR are planned for future work.

We thank the referee for the valuable comments and suggestions on improving the paper. The research is supported by the National Key R\&D Program of China No. 2022YFF0503800, the National Natural Science Foundation of China No. 12173005 and 12350004, and SCOSTEP/PRESTO. J.J. acknowledges the International Space Science Institute Team 474 for fruitful discussions. The SDO/HMI data are courtesy of NASA and the SDO/HMI team. SOHO is a project of international cooperation between ESA and NASA.

\bibliography{sample631}{}
\bibliographystyle{aasjournal}

\end{document}